\documentclass[aps,prl,reprint,superscriptaddress,showpacs,showkeys,preprintnumbers,
nofootinbib,
amsmath,amssymb,
aps,
prl,
]{revtex4-2}

\usepackage{graphicx}
\usepackage{dcolumn}
\usepackage{bm}
\usepackage{mathrsfs}
\usepackage[T1]{fontenc}
\usepackage{scalerel}
\usepackage{siunitx}
\usepackage{physics}
\AtBeginDocument{\RenewCommandCopy\qty\SI}
\DeclareSIUnit\rydberg{\text{Ry}}
\DeclareSIUnit{\au}{a.u.}
\usepackage[normalem]{ulem}
\usepackage[dvipsnames]{xcolor}
\usepackage{hyperref}
\hypersetup{colorlinks=true, linkcolor=blue, citecolor=blue}

\begin{document}

\title{Selective suppression of electronic orders via interlayer coupling in superconducting bilayer nickelate thin films}

\author{Ziao Han\textsuperscript{\#}}
\affiliation{Beijing National Laboratory for Condensed Matter Physics, Institute of Physics, Chinese Academy of Sciences, Beijing 100190, China}
\affiliation{University of Chinese Academy of Sciences, Beijing 100049, China}

\author{Lifen Xiang\textsuperscript{\#}}
\affiliation{Beijing National Laboratory for Condensed Matter Physics, Institute of Physics, Chinese Academy of Sciences, Beijing 100190, China}

\author{Tianren Wang }
\affiliation{Department of Physics, The Chinese University of Hong Kong, Shatin, Hong Kong, China}
\author{Congcong Le}
\affiliation{Hefei National Laboratory, Hefei 230088, China}
\author{Jun Zhan}
\affiliation{Beijing National Laboratory for Condensed Matter Physics, Institute of Physics, Chinese Academy of Sciences, Beijing 100190, China}

\author{Siyi Lei}
\affiliation{Beijing National Laboratory for Condensed Matter Physics, Institute of Physics, Chinese Academy of Sciences, Beijing 100190, China}
\affiliation{University of Chinese Academy of Sciences, Beijing 100049, China}

\author{Sonia Francoual}
\affiliation{Deutsches Elektronen-Synchrotron DESY, Notkestraße 85, 22607 Hamburg, Germany}

\author{Qisi Wang}
\affiliation{Department of Physics, The Chinese University of Hong Kong, Shatin, Hong Kong, China}
\affiliation{State Key Laboratory of Quantum Information Technologies and Materials, The Chinese University of Hong Kong, Shatin, Hong Kong, China}

\author{Jiangping Hu }
\affiliation{Beijing National Laboratory for Condensed Matter Physics, Institute of Physics, Chinese Academy of Sciences, Beijing 100190, China}

\author{Tao Xiang }
\affiliation{Beijing National Laboratory for Condensed Matter Physics, Institute of Physics, Chinese Academy of Sciences, Beijing 100190, China}
\author{Ronny Sutarto}
\affiliation{Canadian Light Source, Saskatoon, SK, Canada}

\author{Xianxin Wu}
\email{xxwu@itp.ac.cn}
\affiliation{Institute of Theoretical Physics, Chinese Academy of Sciences, Beijing, China}

\author{X. J. Zhou }
\email{XJZhou@iphy.ac.cn}
\affiliation{Beijing National Laboratory for Condensed Matter Physics, Institute of Physics, Chinese Academy of Sciences, Beijing 100190, China}
\affiliation{University of Chinese Academy of Sciences, Beijing 100049, China}

\author{Zhihai Zhu }
\email{zzh@iphy.ac.cn}
\affiliation{Beijing National Laboratory for Condensed Matter Physics, Institute of Physics, Chinese Academy of Sciences, Beijing 100190, China}
\affiliation{University of Chinese Academy of Sciences, Beijing 100049, China}

\date{\today}

\begin{abstract}
The discovery of spin-density-wave (SDW) order in bilayer nickelates has intensified interest in its interplay with superconductivity. Unlike cuprates, where doping rapidly suppresses the Néel temperature, the SDW transition temperature ($T_{\mathrm{SDW}}$) in bilayer nickelates is robust against oxygen annealing and even increases under pressure. Here, we combine oxygen annealing with isovalent rare-earth ($A$-site) substitution to effectively apply $c$-axis uniaxial pressure, realizing superconducting bilayer nickelate films with $T_{\mathrm{SDW}}$ suppressed from 150 K to 70 K. Notably, while SDW order is weakened but remains, a second charge-like anisotropy order is completely eliminated in the superconducting state. Polarization-resolved O $K$-edge X-ray absorption and electronic structure calculations show that strengthened interlayer coupling reconstructs the Fermi surface and weakens the SDW. These findings, consistent with a spin–spinless stripe ground state, provide new insight into the mechanism of density wave formation and their interplay with superconductivity.

\end{abstract}

\maketitle

\section{Introduction}

The interplay between magnetism and superconductivity remains one of the central enigmas in the quest to unravel the mechanism of high-temperature superconductivity. In cuprates, this complex relationship is marked by a delicate balance: doping rapidly suppresses antiferromagnetic order, yet the spin-charge stripe phase can persist and even intertwine with superconductivity, with the characteristic temperature tracking $T_c$\cite{keimer_quantum_2015,tranquada_evidence_1995,fradkin_colloquium_2015}. These intricate correlations between magnetism and Cooper pairing have established a vital foundation for exploring emergent quantum materials.    

Recently, bilayer nickelates, most notably La$_3$Ni$_2$O$_7$ and its rare-earth-substituted derivative, have emerged as a platform for high-temperature superconductivity, with $T_c \sim 80$~K under high pressure\cite{sun_signatures_2023,hou_emergence_2023,wang_pressureinduced_2024,zhang_hightemperature_2024} and ambient-pressure superconductivity above 40~K achieved in epitaxial thin films\cite{ko_signatures_2025a,xiang_stabilizing_2026,liu_superconductivity_2025,zhou_ambientpressure_2025a,zhou_superconductivity_2026,hao_superconductivity_2025}. Substituting La with smaller rare-earth ions such as Pr, Sm, and Nd further raises $T_c$ toward $\sim$100~K\cite{wang_bulk_2024,li_bulk_2026,qiu_hightemperature_2026}. Unlike the cuprates, which can be described effectively within a single-band (Cu $d_{x^2-y^2}$) framework, La$_3$Ni$_2$O$_7$ and its variants adopt a bilayer structure with multiorbital character (Ni $d_{x^2-y^2}$ and $d_{z^2}$). In particular, the Ni $d_{z^2}$ orbitals hybridize through the apical oxygens between the two NiO$_2$ layers, generating strong interlayer coupling and giving rise to an interlayer bonding $\gamma$ band, the position of which relative to the Fermi level remains under active debate\cite{li_angleresolved_2025,wang_electronic_2025,nie_superconductivity_2026,yang_orbitaldependent_2024,sun_observation_2025,luo_bilayer_2023,zhang_structural_2024,ryee_superconductivity_2025,qu_bilayer_2024,yang_interlayer_2023,jiang_hightemperature_2024}. 

A spin-density wave (SDW) order has now been observed in the non-superconducting regime of bilayer nickelates\cite{ren_resolving_2025a,gupta_anisotropic_2025,yashima_microscopic_2025,plokhikh_unraveling_2025,chen_electronic_2024,chen_evidence_2024}, accompanied by strong electronic correlations and the emergence of a partial gap\cite{liu_electronic_2024,wang_normal_2024}. The magnetic structure of this SDW remains under debate\cite{ren_resolving_2025a, gupta_anisotropic_2025,khasanov_pressureenhanced_2025,yashima_microscopic_2025,chen_nature_2026}. Recent Nuclear Quadrupole Resonance (NQR)\cite{yashima_microscopic_2025} and neutron scattering experiments\cite{chen_nature_2026}, however, provide microscopic evidence for a spin–spinless stripe order, in which the spin and spinless sublattices carry different degrees of freedom. In this framework, the SDW and the charge-like anisotropy order observed in resonant elastic X-ray scattering (REXS) naturally map onto the spin and spinless sublattices, respectively\cite{ren_resolving_2025a}. Muon-spin rotation ($\mu$SR) experiments reveal that the SDW transition temperature ($T_{\mathrm{SDW}}$) in La$_3$Ni$_2$O$_{7-\delta}$ increases under pressure, while the second density wave transition temperature decreases\cite{khasanov_pressureenhanced_2025}, indicating that the two orders respond differently to pressure. In thin films, oxygen annealing reduces the SDW peak intensity in REXS experiments but leaves its $T_{\mathrm{SDW}}$ largely unchanged\cite{ren_resolving_2025a,chen_3d_z2_2026,zhang_interlayer_2026,li_interlayer_2026,gupta_anisotropic_2025}. The magnetic structure and the nature of these ordered states remain unclear, as does their relationship to superconductivity.

\begin{figure*}[htb] \centering
\includegraphics[width=0.95\textwidth]{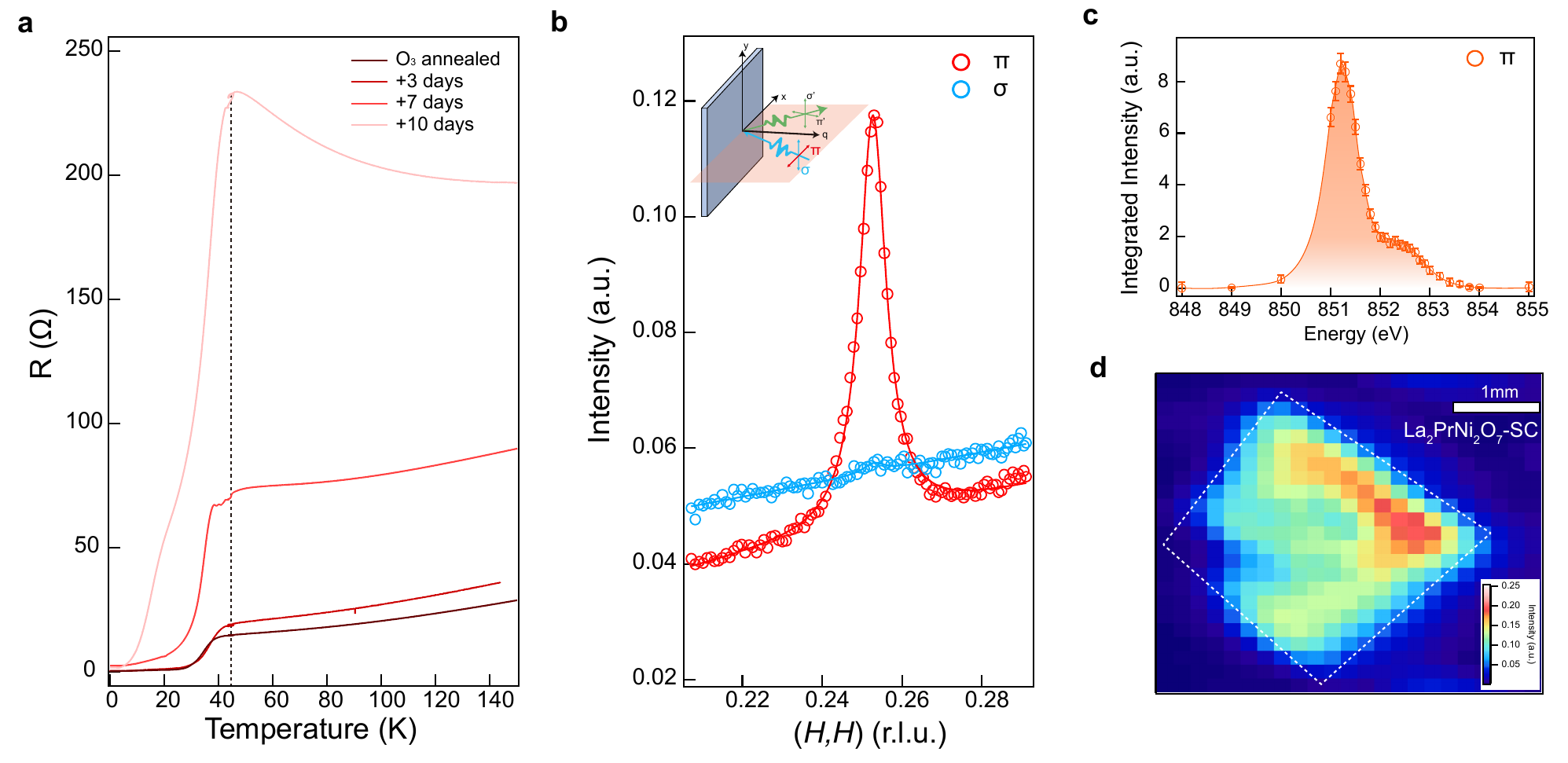}
\caption{\textbf{SDW order and transport characterization of superconducting La$_2$PrNi$_2$O$_{7-\delta}$ thin films on SrLaAlO$_4$.}
(a) $R(T)$ curves measured at 0, 3, 7, and 10 days after ozone annealing; the sample for the REXS measurements was cleaved from the same film and packaged under identical conditions.
(b) Resonant elastic X-ray scattering (REXS) scans across the SDW wavevector $\mathbf{Q} = (0.25, 0.25, 1.9)$ measured at 20~K with both $\pi$- and $\sigma$-polarized incident X-rays tuned to the Ni $L_3$ absorption edge ($\sim$852.3~eV). The solid lines are Lorentzian fits.
(c) Integrated SDW peak intensity versus incident photon energy, reaching a maximum near 851.3~eV, consistent with the Ni $L_3$ resonance.
(d) Real-space mesh scan of the maximum SDW peak intensity.}
\label{fig1}
\end{figure*}

Here, we address these questions by systematically tuning the interlayer coupling in bilayer nickelate films through a combination of $A$-site rare-earth substitution and ozone annealing. Using REXS, we find that the SDW and charge-like anisotropy orders respond differently: $T_{\mathrm{SDW}}$ decreases from 150~K to 70~K and the SDW order persists, whereas the charge-like anisotropy order is completely eliminated. $A$-site substitution compresses the $c$-axis, while ozone annealing fills interlayer oxygen vacancies and restores the $3d_{z^2}$--$2p_z$--$3d_{z^2}$ hybridization channel—two distinct routes that both enhance interlayer coupling. Density functional theory calculations indicate that this enhanced interlayer coupling reconstructs the Fermi surface and suppresses spin fluctuations near the SDW wavevector. This selective suppression demonstrates that the two orders are not governed by a single energy scale, providing a new constraint on the origins of these orders and on how they relate to superconductivity in bilayer nickelates.

\section{Results and discussion}

 We observe well-defined SDW peaks in bilayer nickelate thin films that also exhibit robust superconductivity. REXS measurements were performed on freshly prepared films within a three-day window under controlled conditions, and the stability of the superconducting state was verified by the test shown in Fig.~\ref{fig1}(a). We present in Fig.~\ref{fig1}(b) representative rocking curves at the wavevector $\mathbf{Q}=(0.25,0.25,1.9)$, measured on a La$_2$PrNi$_2$O$_7$ (LPNO) film using both $\pi$- and $\sigma$-polarized incident photons at the Ni $L_3$ absorption edge. Figure
 ~\ref{fig1}(c) displays the energy profile of the reflection shown in Fig.~\ref{fig1}(b) for $\pi$-polarized incident photons, confirming the electronic origin of the signal in Ni $d$ orbitals. These results are consistent with previous reports on related nickelate systems and further confirm the robustness of the superlattice structure under varying sample conditions\cite{ren_resolving_2025a,gupta_anisotropic_2025}. To assess the spatial distribution of the SDW phase, we performed real-space mesh scans of the peak intensity. As shown in Fig.~\ref{fig1}(d), nonzero SDW peak intensity is detected at every probed position, with no regions lacking SDW signal within the instrumental resolution limit ($\sim$450 $\mu$m(L) $\times$ $\sim$450 $\mu$m(W)). The modest point-to-point intensity variation arises from extrinsic factors, such as silver paint used for grounding at the film corners and the gold electrode pattern on the sample surface, rather than intrinsic electronic inhomogeneity.
 
\begin{figure*}[htb] \centering
\includegraphics[width=0.95\textwidth]{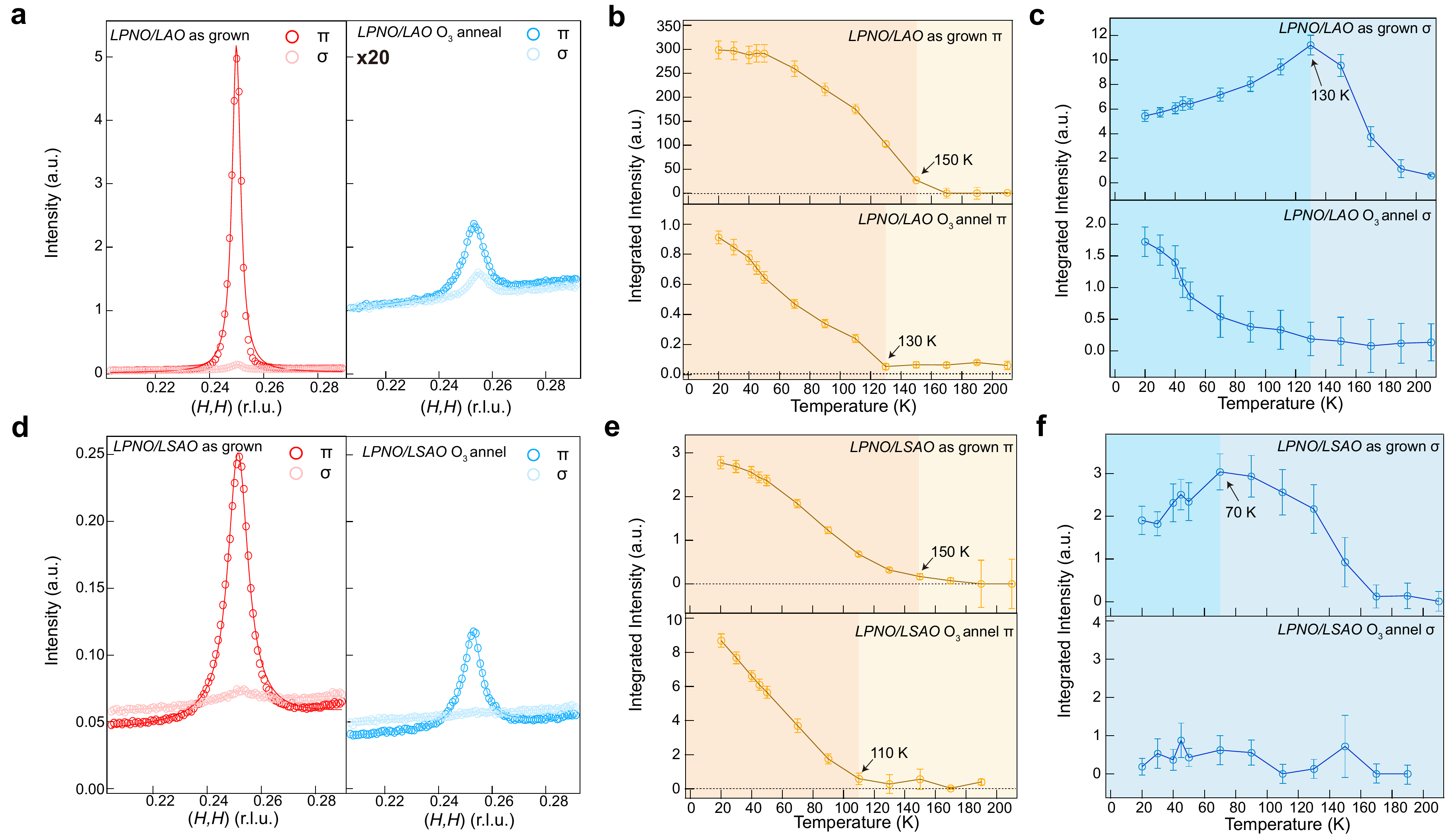}
\caption{\textbf{SDW peak and temperature dependence under oxygen-content tuning in La$_2$PrNi$_2$O$_{7-\delta}$ thin films.}
(a) REXS scans across the SDW wavevector $\mathbf{Q} = (0.25, 0.25, 1.9)$ measured at 20 K with both $\pi$- and $\sigma$-polarized incident X-rays at the Ni $L_3$ edge. Left: as-grown LPNO thin film ($\sim$26 nm) grown on LaAlO$_3$ (LAO). Right: ozone-annealed LPNO thin film ($\sim$28 nm) grown on LAO. The solid lines are Lorentzian fits to the SDW peak. (b) Temperature dependence of the integrated SDW peak intensity measured with $\pi$-polarization for the as-grown (top) and ozone-annealed (bottom) LPNO/LAO films. (c) Temperature dependence of the integrated SDW peak intensity measured with $\sigma$-polarization for the same LPNO/LAO films. (d) REXS scans across $\mathbf{Q} = (0.25, 0.25, 1.9)$ measured at 20 K with $\pi$- and $\sigma$-polarization. Left: as-grown LPNO thin film ($\sim$7 nm) grown on SrLaAlO$_4$ (SLAO). Right: ozone-annealed superconducting LPNO thin film ($\sim$8 nm) grown on SLAO. (e) Temperature dependence of the integrated SDW peak intensity measured with $\pi$-polarization for the as-grown (top) and ozone-annealed superconducting (bottom) LPNO/SLAO films. (f) Temperature dependence of the integrated SDW peak intensity measured with $\sigma$-polarization for the same LPNO/SLAO films.}
\label{fig2}
\end{figure*}

Oxygen stoichiometry is a critical factor governing the superconducting and electronic properties of bilayer nickelate thin films\cite{ren_resolving_2025a,xiang_stabilizing_2026}. To elucidate its effect on SDW order, we compare LPNO films grown on LaAlO$_3$ (LAO) before and after ozone annealing. Figure~\ref{fig2}(a) presents $\pi$- and $\sigma$-polarized REXS scans at $\mathbf{Q} = (0.25, 0.25, 1.9)$ and 20 K for both as-grown and ozone-annealed samples. In each case, the $\pi$-channel peak intensity greatly exceeds that of the $\sigma$-channel, consistent with a predominantly magnetic origin of the reflection. Ozone annealing leads to a pronounced suppression of the resonant peak intensity in both polarization channels. To enhance the visibility of the weak peaks, the right panel of Fig.~\ref{fig2}(a) displays the data for the annealed film with the vertical axis scaled by a factor of 20 to highlight the residual signal.

\begin{figure*}[htb] \centering
\includegraphics[width=0.95\textwidth]{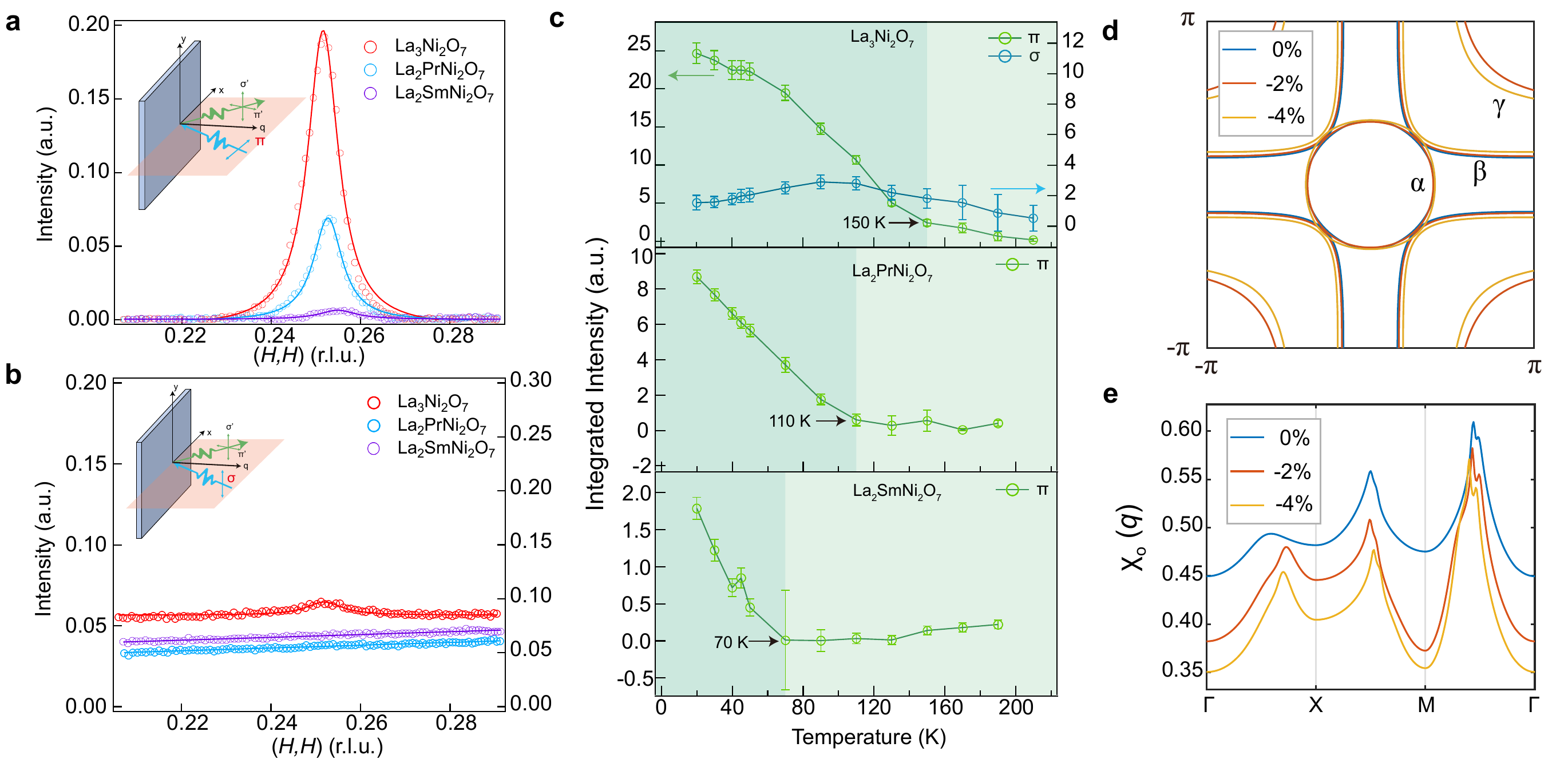}
\caption{\textbf{Comparison of SDW order across La$_3$Ni$_2$O$_7$, La$_2$PrNi$_2$O$_7$, and La$_2$SmNi$_2$O$_7$ thin films.}
(a) REXS scans across the SDW wavevector $\mathbf{Q} = (0.25, 0.25, 1.9)$ measured at 20 K with $\pi$-polarized incident X-rays at the Ni $L_3$ edge for superconducting La$_3$Ni$_2$O$_{7-\delta}$, La$_2$PrNi$_2$O$_{7-\delta}$, and La$_2$SmNi$_2$O$_{7-\delta}$ thin films. The solid lines are Lorentzian fits to the SDW peak. (b) Corresponding REXS scans measured with $\sigma$-polarization for the same three films. (c) Temperature dependence of the integrated SDW peak intensity for La$_3$Ni$_2$O$_{7-\delta}$ (top), La$_2$PrNi$_2$O$_{7-\delta}$ (middle), and La$_2$SmNi$_2$O$_{7-\delta}$ (bottom). For La$_3$Ni$_2$O$_{7-\delta}$, both $\pi$-polarization (green) and $\sigma$-polarization (blue) data are shown. (d) Calculated Fermi surfaces in the $k_x$--$k_y$ plane ($-\pi$ to $\pi$) for undoped, Pr-doped, and Sm-doped compositions, corresponding to the colors in (e). (e) Calculated electronic susceptibility $\chi_0(\mathbf{q})$ along the high-symmetry path $\Gamma$--X--M--$\Gamma$ for the same three compositions as in (d). The labels 0\%, $-$2\%, and $-$4\% denote the relative change in $A$-site ionic diameter with respect to La$^{3+}$.}
\label{fig3}
\end{figure*}

The temperature dependence of the integrated peak intensity highlights the strong influence of oxygen stoichiometry on the SDW and charge-like anisotropy order. The $\sigma$-channel signal contains a charge-like anisotropy order\cite{ren_resolving_2025a} and shows a hallmark non-monotonic temperature profile. In the $\pi$ channel, the as-grown LPNO/LAO film exhibits an SDW transition at $T_{\mathrm{SDW}} \sim 150$~K (Fig.~\ref{fig2}(b), top), which shifts down to $\sim$130~K after ozone annealing (Fig.~\ref{fig2}(b), bottom). The $\sigma$ channel exhibits a qualitatively different response. In the as-grown film, the non-monotonic temperature profile previously reported for La$_3$Ni$_2$O$_7$\cite{ren_resolving_2025a} is observed: the intensity increases upon cooling but drops sharply below 110~K. After ozone annealing, however, this non-monotonic feature disappears; instead, the $\sigma$-channel intensity decays monotonically with increasing temperature, indicating the suppression of the charge-like anisotropy order.

We next examine thinner LPNO films ($\sim$8 nm) grown on SLAO substrates. In contrast to the thicker LPNO films on LAO, which remain non-superconducting, these SLAO-grown films become superconducting after ozone annealing. Figure~\ref{fig2}(d) presents a direct comparison for the SLAO-grown films. In the superconducting sample, the SDW peak is strongly suppressed in the $\pi$ channel. The temperature-dependent SDW peak intensity, shown in Figs.~\ref{fig2}(e) and \ref{fig2}(f), provides further insight into this behavior. In the $\pi$ channel, $T_{\mathrm{SDW}}$ shifts from approximately 150~K in the non-superconducting film to about 110~K in the superconducting film, as shown in Fig.~\ref{fig2}(e). Moreover, upon warming from 20 K, the non-superconducting film shows a weak tendency toward saturation at low temperatures followed by a gradual decline, while the superconducting film shows a nearly linear decrease across the entire range. In the $\sigma$ channel, the non-superconducting film retains the non-monotonic profile, with a turning point near 70~K (Fig.~\ref{fig2}(f), top), whereas the superconducting film shows no discernible peak at any measured temperature (Fig.~\ref{fig2}(f), bottom).

Beyond oxygen content, $A$-site rare-earth substitution provides another important pathway to tune interlayer coupling\cite{li_bulk_2026,wang_bulk_2024,pan_effect_2024}.
Figure~\ref{fig3}(a) compares $\pi$-polarized REXS scans at 20 K for three superconducting films grown on SLAO: La$_3$Ni$_2$O$_{7-\delta}$ (LNO), La$_2$PrNi$_2$O$_{7-\delta}$ (LPNO), and La$_2$SmNi$_2$O$_{7-\delta}$ (LSNO). The SDW peak intensity decreases progressively from the undoped LNO to Pr-doped LPNO and further to Sm-doped LSNO. In the $\sigma$ channel, shown in Fig.~\ref{fig3}(b), only the parent LNO film retains a weak but discernible peak, while the Pr- and Sm-doped samples exhibit no resolvable signal above the background. The two channels therefore respond differently to A-site substitution as well: the $\pi$-channel SDW is progressively weakened but persists, whereas the $\sigma$-channel charge-like order is completely eliminated upon substitution with Pr or Sm. This systematic evolution with rare-earth substitution is further reflected in the temperature dependence of the SDW order. As shown in Fig.~\ref{fig3}(c), the SDW transition temperature decreases with increasing chemical pressure: the $\pi$-channel integrated intensity drops with increasing temperature, with characteristic $T_{\mathrm{SDW}}$ values of $150$~K for LNO, 110 K for LPNO, and 70 K for LSNO. In contrast, in bulk samples, Pr substitution yields only a minimal change in the SDW transition temperature ($T_N$ rises slightly from $\sim$156.5 K to $\sim$161.4 K)\cite{chen_prdoping_2025,plokhikh_unraveling_2025,khasanov_pressureeffect_2025}.

\begin{figure*}[htb] \centering
\includegraphics[width=1\textwidth]{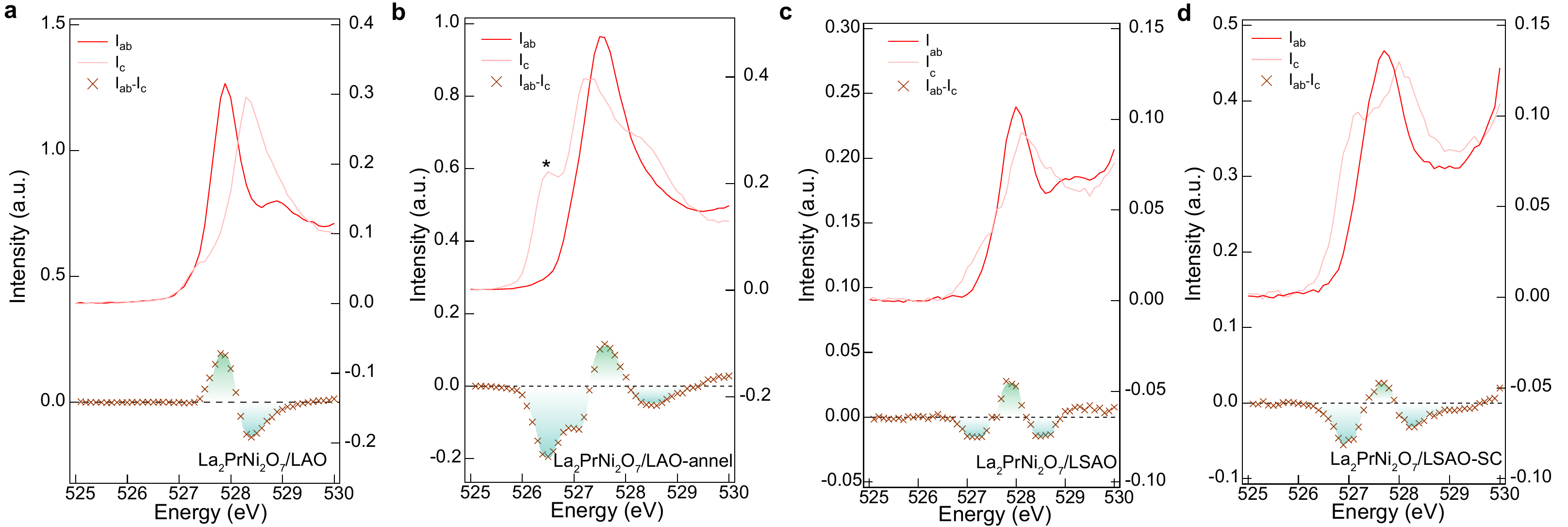}
\caption{\textbf{O $K$-edge pre-edge XAS and XLD of LPNO thin films.}
(a)--(d) O $K$-edge X-ray absorption spectra (XAS) measured at 20 K showing the pre-edge region. In-plane ($I_{ab}$, red) and out-of-plane ($I_c$, pink) polarized spectra together with the X-ray linear dichroism (XLD $= I_{ab} - I_c$, crosses) are presented for: (a) as-grown La$_2$PrNi$_2$O$_{7-\delta}$ (LPNO) thin film grown on LaAlO$_3$ (LAO). (b) Ozone-annealed LPNO thin film grown on LAO; the star denotes a peak feature of unknown origin, likely associated with a phase introduced by ozone annealing. (c) As-grown LPNO thin film grown on SrLaAlO$_4$ (SLAO). (d) Ozone-annealed superconducting LPNO thin film grown on SLAO.}
\label{fig4}
\end{figure*}

To understand how $A$-site substitution suppresses the SDW order, we performed theoretical calculations of the electronic structure and spin susceptibility. The partial substitution of La with smaller rare-earth ions is modeled as a uniaxial compressive strain along the $c$-axis. This strain enhances the interlayer coupling, expanding the $\alpha$ pocket while contracting the $\beta$ pocket [Fig.~\ref{fig3}(d)]. As a result, the $\alpha$--$
\beta$ inter-pocket scattering is reduced, which leads to a decrease in the odd-channel magnetic susceptibility near $(\pi/2, \pi/2)$ [Fig.~\ref{fig3}(e)]. This reduction in spin fluctuations suppresses the SDW instability and accounts for the progressive weakening of the SDW state from La$_3$Ni$_2$O$_7$ to La$_2$PrNi$_2$O$_7$ and La$_2$SmNi$_2$O$_7$.

Combined with the oxygen-content results in Fig.~\ref{fig2}, our observations show that the two orders respond differently to the same tuning: the SDW weakens progressively as superconductivity is optimized, whereas the charge-like anisotropy order is completely eliminated. Notably, the temperature dependence of SDW integrated intensity shows no anomaly at $T_c$, indicating that the SDW order and superconductivity do not exhibit the direct competition of the type that exists, for example, between charge order and superconductivity in cuprates\cite{chang_direct_2012,ghiringhelli_longrange_2012}. The charge-like anisotropy order observed in the $\sigma$-channel is even more sensitive: its characteristic non-monotonic temperature profile in the as-grown films--with turning points near 130 K on LAO and 70 K on SLAO--is completely suppressed once superconductivity is established. To exclude a structural origin for the non-monotonic anomaly, we performed hard X-ray diffraction on the LNO film, which shows no signature of a structural transition in this temperature range (see Supplementary Note 6 for further details). We thus attribute the anomaly to an electronic instability, possibly linked to Fermi-surface reconstruction\cite{liu_electronic_2024,liu_evidence_2023}.

 We performed polarization-resolved XAS at the O $K$-edge to investigate the impact of oxygen annealing on the interlayer hybridization channel. Within the Zaanen-Sawatzky-Allen (ZSA) charge-transfer framework\cite{zaanen_band_1985}, the O $K$-edge pre-edge is sensitive to ligand-hole states hybridized with Ni $3d$ orbitals. As shown in Fig.~\ref{fig4}, the in-plane spectra $I_{ab}$ show a sharp peak near 528~eV and a weaker feature near 529~eV. Following ozone annealing, the weak feature disappears while the sharp peak remains essentially unchanged, indicating that the in-plane $d_{x^2-y^2}$–$p_{x,y}$ conduction channel is robust and largely insensitive to oxygen content\cite{zhang_interlayer_2026}. In contrast, the out-of-plane spectra, $I_{c}$, comprise two peaks: a low-energy peak (LE, $\sim$527.1--527.2~eV) and a high-energy peak (HE, $\sim$528.0--528.3~eV), which are more sensitive to oxygen stoichiometry. 

 Figure~\ref{fig4}(a,b) presents polarization-resolved XAS measurements of the non-superconducting LPNO/LAO film ($\sim$30 nm). Our calculations for the stoichiometric compound indicate that the low-energy spectral weight along the out-of-plane direction primarily originates from the interlayer apical oxygens (see Supplementary Note 7). In contrast, the X-ray linear dichroism (XLD) of the as-grown film lacks the corresponding low-energy dip, which is restored after ozone annealing. The disappearance of the dip, therefore, is consistent with a significant concentration of interlayer oxygen vacancies in the as-grown film, in line with previous atomic-scale imaging in La$_3$Ni$_2$O$_{7-\delta}$\cite{dong_visualization_2024}. Although Pr substitution suppresses these vacancies in bulk samples\cite{dong_interstitial_2025}, a substantial number of interlayer oxygen vacancies persist in our thick LPNO/LAO films. Ozone annealing fills these oxygen vacancies and restores the low-energy spectral weight, bringing the spectrum closer to the stoichiometric calculation. Figures~\ref{fig4}(c) and \ref{fig4}(d) display the corresponding XAS for LPNO films with a thickness of $\sim$8 nm grown on SLAO. In contrast to the $\sim$30 nm LPNO/LAO film, both non-superconducting and ozone-annealed superconducting LPNO/SLAO films exhibit the low-energy dip, indicating that interlayer oxygen vacancies are largely suppressed. Notably, the low-energy spectral weight is further enhanced in the superconducting film relative to the non-superconducting sample, indicating increased interlayer hybridization upon ozone annealing. These observations confirm that ozone annealing restores the $3d_{z^2}$--$2p_z$--$3d_{z^2}$ interlayer hybridization channel.

The results above can be understood within a unified framework in which interlayer coupling plays a central role in controlling SDW order in bilayer nickelates. $A$-site substitution effectively compresses the $c$-axis, imposing uniaxial pressure that enhances interlayer interactions. Ozone annealing achieves a similar effect by filling interlayer apical oxygen vacancies and restoring the $3d_{z^2}$--$2p_z$--$3d_{z^2}$ hybridization channel, as confirmed by our XAS measurements. Both approaches lead to increased spectral weight of the $\gamma$ band near the Fermi level, as evidenced by the spectral weight transfer observed in our polarization-resolved XAS using the band assignment of Ref.~\cite{ren_resolving_2025a}. 

Our XAS measurements show that ozone annealing restores the $3d_{z^2}$--$2p_z$--$3d_{z^2}$ hybridization channel. Together with our calculations demonstrating that A-site substitution enhances this channel via $c$-axis compression, these results identify interlayer coupling as the central parameter controlling the two electronic orders. While the microscopic origin of SDW order in bilayer nickelates remains under debate \cite{ren_resolving_2025a,gupta_anisotropic_2025,yashima_microscopic_2025,plokhikh_unraveling_2025,chen_electronic_2024,chen_evidence_2024,luo_bilayer_2023,mei_itinerant_2026,lechermann_electronic_2023,christiansson_correlated_2023,sakakibara_possible_2024,chen_electronic_2024,au-yeung_oxygencentred_2026}, this spectral-weight transfer can be interpreted from both itinerant and strong-correlation perspectives. In the itinerant view, increased $\gamma$-band weight at the Fermi level alters the nesting conditions between the $\alpha$ and $\beta$ pockets, thereby weakening the SDW instability. From the perspective of strong correlations, the same transfer signifies the delocalization of Ni $d_{z^2}$ orbitals as coherent interlayer hybridization develops\cite{zhang_interlayer_2026,shilenko_correlated_2023,pan_superconductivity_2026}; since static SDW order is stabilized by local moments on the Ni sites, this delocalization undermines the moments and suppresses long-range order. In both scenarios, enhanced interlayer coupling destabilizes SDW order. This explains the suppression of the $\pi$-channel SDW, but not the fate of the $\sigma$-channel charge-like anisotropy order. To understand this selective suppression, we return to the spin–spinless stripe framework\cite{ren_resolving_2025a,chen_nature_2026,yashima_microscopic_2025}, within which the charge-like anisotropy order is associated with a magnetic quadrupole on the spinless sublattices. Such a quadrupole creates an anisotropic local environment that couples to neighboring dipole moments and stabilizes the static SDW order by locking the orientation of adjacent moments. Eliminating this anisotropic environment weakens SDW correlations, which is consistent with the reduced SDW order observed in the superconducting films.

\section{Conclusion}
In summary, our study demonstrates that $T_{\mathrm{SDW}}$ in bilayer nickelates can be effectively tuned by enhancing interlayer coupling through rare-earth substitution and oxygen annealing. This controlled tuning provides a route to disentangle the mechanisms governing multiple electronic orders. We find that the suppression of the $\pi$-channel SDW order can be consistently explained by an itinerant, Fermi surface-driven scenario, reflecting the role of inter-pocket nesting. In contrast, the complete disappearance of the $\sigma$-channel charge-like anisotropy order under the same conditions cannot be accounted for by band theory alone, and instead points to the essential role of electronic correlations. These results provide a new constraint on the origins of these density waves and on their relationship to superconductivity in bilayer nickelate films.

\section{Methods}

\subsection{Sample preparation}

Thin films of La$_2$PrNi$_2$O$_{7-\delta}$ (LPNO), La$_3$Ni$_2$O$_{7-\delta}$ (LNO), and La$_2$SmNi$_2$O$_{7-\delta}$ (LSNO) were grown on (001)-oriented LaAlO$_3$ (LAO) and SrLaAlO$_4$ (SLAO) substrates using pulsed laser deposition with a 248-nm KrF excimer laser. A stoichiometric polycrystalline target was ablated at a substrate temperature of 680$\,^{\circ}$C under an oxygen partial pressure of 150~mTorr. Film thicknesses, calibrated by X-ray reflectivity, ranged from approximately 5.5~nm to 28~nm, depending on the growth duration. After deposition, the films were cooled to room temperature at a rate of 5$\,^{\circ}$C~min$^{-1}$ in an oxygen partial pressure of 150~mTorr. To achieve the superconducting state, selected films were subjected to an ozone-assisted annealing procedure following the oxygen recycling protocol described in Ref.\cite{xiang_stabilizing_2026}.

\subsection{XAS and REXS measurements}

The X-ray absorption spectroscopy (XAS) and resonant elastic X-ray scattering (REXS) measurements at the Ni $L$-edge and O $K$-edge were performed at the REIXS (10ID-2) beamline of the Canadian Light Source (CLS)\cite{hawthorn_invacuum_2011}. The beamline delivers a photon flux of approximately $5\times 10^{11}$~photons~s$^{-1}$ with an energy resolution $\Delta E/E \sim 2\times 10^{-4}$ in this photon energy range. An in-vacuum four-circle diffractometer housed the samples under ultra-high vacuum conditions below $5\times 10^{-10}$~mbar. The XAS spectra were collected in both total electron yield (TEY) and total fluorescence yield (TFY) modes at 20~K with linear vertical ($\sigma$) and horizontal ($\pi$) light polarizations. The spectra were normalized to the incident photon flux. The REXS spectra were obtained by performing momentum scans across the wavevector of interest at selected photon energies and temperatures using a silicon drift detector (SDD), and the peak intensities were extracted from Lorentzian fits to the data on a linear background. The hard X-ray diffraction was performed close to the Ni $K$-edge at beamline P09 at PETRA III in the first experimental station equipped with a six-circle diffractometer. The sample was mounted at the tip of the cold finger of a closed cycle cryostat. The temperature dependent measurements were performed in the vertical scattering geometry without any polarization analyzer in front of the detector.

\subsection{Band structure calculations }
The band structures and O $K$-edge X-ray absorption spectra were calculated using the projector augmented wave (PAW) method, as implemented in the Vienna ab initio simulation package (VASP). The exchange-correlation energy was treated using the Perdew-Burke-Ernzerhof (PBE) functional \cite{perdew_generalized_1996}, and a plane-wave kinetic-energy cutoff of 550 eV was employed. For the band structure calculations, lattice parameters of a = b = 3.75 \AA~ and c = 20.865 \AA~ were adopted. The partial substitution of La with smaller rare-earth ions was modeled by systematically reducing the $c$-axis lattice constant. To construct an effective model of the low-energy electronic structure, we employed the Wannier90 package to calculate maximally localized Wannier functions spanning the Ni $d_{x^2-y^2}$ and $d_{z^2}$ orbitals, which were subsequently used to evaluate the magnetic susceptibility.
To simulate the O $K$-edge X-ray absorption spectra, a 2$\times$2$\times$1 supercell was used, with the Brillouin zone sampled by a $\Gamma$-centered 6$\times$6$\times$2 k-point mesh based on the Monkhorst-Pack scheme. The planar, inner-apical, and outer-apical oxygen sites were treated separately. For each inequivalent oxygen site, one O 1s electron was removed and promoted to the unoccupied manifold utilizing the excited-electron and core-hole approach, ensuring the overall charge neutrality of the supercell was maintained.\\

\noindent\textbf{Acknowledgments}\\
We thank George Sawatzky for his very insightful comments on our manuscript. This work was supported in part by the National Key Research and Development Program of China (Grant No. 2022YFA1403900 and 2021YFA1401800), the National Natural Science Foundation of China (Grant No. 12494593), the Strategic Priority Research Program (B) of the Chinese Academy of Sciences (Grant No. XDB25000000), and the Synergetic Extreme Condition User Facility (SECUF). T.W. and Q.W. are supported by the Research Grants Council of Hong Kong (CUHK 24306223, CUHK 14305926). Part of the research described in this paper was performed at the Canadian Light Source, a national research facility of the University of Saskatchewan, which is supported by the Canada Foundation for Innovation (CFI), the Natural Sciences and Engineering Research Council (NSERC), the Canadian Institutes of Health Research (CIHR), the Government of Saskatchewan, and the University of Saskatchewan. We acknowledge DESY (Hamburg, Germany), a member of the Helmholtz Association HGF, for the provision of experimental facilities. Parts of this research were carried out at beamline P09 at the PETRA III synchrotron. Beamtime was allocated for proposal I-20251220.\\

\noindent\textbf{Author contributions}\\
Z.H. and L.X. contributed equally to this work. Z.H.Z. and X.J.Z. conceived and supervised the project. Z.H. and L.X. grew and characterized the films with assistance from S.L. Z.H., T.W., Q.W., R.S., and Z.H.Z. performed the XAS and REXS experiments at the REIXS (10ID-2) beamline of the Canadian Light Source (CLS). T.W., S.F., Z.H., L.X., Q.W., and Z.H.Z. conducted the hard X-ray diffraction measurements at the P09 beamline of DESY. C.L., J.Z., X.W., J.H., and T.X. contributed to the theoretical calculations and analysis. Z.H. and Z.H.Z. wrote the manuscript with input from all authors.\\

\noindent\textbf{Data availability} \\
Data supporting the findings of this study are available from the corresponding authors on a reasonable request.\\

\bibliography{reference}

@article{keimer_quantum_2015,
  title = {From Quantum Matter to High-Temperature Superconductivity in Copper Oxides},
  author = {Keimer, B. and Kivelson, S. A. and Norman, M. R. and Uchida, S. and Zaanen, J.},
  year = 2015,
  month = feb,
  journal = {Nature},
  volume = {518},
  number = {7538},
  pages = {179--186},
  issn = {0028-0836, 1476-4687},
  doi = {10.1038/nature14165},
  url = {https://www.nature.com/articles/nature14165},
  urldate = {2026-07-10},
  langid = {english}
}

@article{tranquada_evidence_1995,
  title = {Evidence for Stripe Correlations of Spins and Holes in Copper Oxide Superconductors},
  author = {Tranquada, J. M. and Sternlieb, B. J. and Axe, J. D. and Nakamura, Y. and Uchida, S.},
  year = 1995,
  month = jun,
  journal = {Nature},
  volume = {375},
  number = {6532},
  pages = {561--563},
  issn = {0028-0836, 1476-4687},
  doi = {10.1038/375561a0},
  url = {https://www.nature.com/articles/375561a0},
  urldate = {2026-07-10},
  copyright = {http://www.springer.com/tdm},
  langid = {english}
}

@article{sun_signatures_2023,
  title = {Signatures of Superconductivity near 80 {{K}} in a Nickelate under High Pressure},
  author = {Sun, Hualei and Huo, Mengwu and Hu, Xunwu and Li, Jingyuan and Liu, Zengjia and Han, Yifeng and Tang, Lingyun and Mao, Zhongquan and Yang, Pengtao and Wang, Bosen and Cheng, Jinguang and Yao, Dao-Xin and Zhang, Guang-Ming and Wang, Meng},
  year = 2023,
  month = sep,
  journal = {Nature},
  volume = {621},
  number = {7979},
  pages = {493--498},
  issn = {0028-0836, 1476-4687},
  doi = {10.1038/s41586-023-06408-7},
  url = {https://www.nature.com/articles/s41586-023-06408-7},
  urldate = {2026-04-01},
  langid = {english}
}

@article{hou_emergence_2023,
  title = {{Emergence of HighTemperature Superconducting Phase in Pressurized La$_{3}$Ni$_{2}$O$_{7}$ Crystals}},
  author = {Hou, Jun and Yang, Peng-Tao and Liu, Zi-Yi and Li, Jing-Yuan and Shan, Peng-Fei and Ma, Liang and Wang, Gang and Wang, Ning-Ning and Guo, Hai-Zhong and Sun, Jian-Ping and Uwatoko, Yoshiya and Wang, Meng and Zhang, Guang-Ming and Wang, Bo-Sen and Cheng, Jin-Guang},
  year = {2023},
  month = {oct},
  journal = {Chin. Phys. Lett.},
  volume = {40},
  number = {11},
  pages = {117302},
  publisher = {{Chinese Physical Society and IOP Publishing Ltd}},
  doi = {10.1088/0256-307X/40/11/117302},
  url = {https://iopscience.iop.org/article/10.1088/0256-307X/40/11/117302},
}

@article{zhang_hightemperature_2024,
  title = {{High Temperature Superconductivity with Zero Resistance and StrangeMetal Behaviour in {La$_{3}$Ni$_{2}$O$_{7-\delta}$}}},
  author = {Zhang, Ya-Nan and Su, Da-Jun and Huang, Ya-Nen and Shan, Zhao-Yang and Sun, Hua-Lei and Huo, Meng-Wu and Ye, Kai-Xin and Zhang, Jia-Wen and Yang, Zi-Han and Xu, Yong-Kang and Su, Yi and Li, Rui and Smidman, Michael and Wang, Meng and Jiao, Lin and Yuan, Hui-Qiu},
  year = 2024,
  month = aug,
  journal = {Nat. Phys.},
  volume = {20},
  number = {8},
  pages = {1269--1273},
  issn = {1745-2473, 1745-2481},
  doi = {10.1038/s41567-024-02515-y},
  url = {https://www.nature.com/articles/s41567-024-02515-y},
  urldate = {2026-03-10},
  langid = {english},
 
}

@article{wang_pressureinduced_2024,
 title = {{PressureInduced Superconductivity In Polycrystalline La$_{2}$PrNi$_{2}$O$_{7-\delta}$}},
	volume = {14},
	url = {https://link.aps.org/doi/10.1103/PhysRevX.14.011040},
	doi = {10.1103/PhysRevX.14.011040},
	number = {1},
	journal = {Phys. Rev. X},
	author = {Wang, G. and Wang, N. N. and Shen, X. L. and Hou, J. and Ma, L. and Shi, L. F. and Ren, Z. A. and Gu, Y. D. and Ma, H. M. and Yang, P. T. and Liu, Z. Y. and Guo, H. Z. and Sun, J. P. and Zhang, G. M. and Calder, S. and Yan, J.Q. and Wang, B. S. and Uwatoko, Y. and Cheng, J.G.},
	month = mar,
	year = {2024},
	pages = {011040},
}

@article{li_bulk_2026,
  title = {{Bulk Superconductivity up to 96 {{K}} in Pressurized Nickelate Single Crystals}},
  author = {Li, Fei-Yu and Xing, Zhen-Fang and Peng, Di and Dou, Jie and Guo, Ning and Ma, Liang and Zhang, Yu-Lin and Wang, Ling-Zhen and Luo, Jun and Yang, Jie and Zhang, Jian and Chang, Tie-Yan and Chen, Yu-Sheng and Cai, Wei-Zhao and Cheng, Jin-Guang and Wang, Yu-Zhu and Liu, Yu-Xin and Luo, Tao and Hirao, Naohisa and Matsuoka, Takahiro and Kadobayashi, Hirokazu and Zeng, Zhi-Dan and Zheng, Qiang and Zhou, Rui and Zeng, Qiao-Shi and Tao, Xu-Tang and Zhang, Jun-Jie},
  year = 2026,
  month = jan,
  journal = {Nature},
  volume = {649},
  number = {8098},
  pages = {871--878},
  issn = {0028-0836, 1476-4687},
  doi = {10.1038/s41586-025-09954-4},
  url = {https://www.nature.com/articles/s41586-025-09954-4},
  urldate = {2026-04-01},
  langid = {english},
  
}

@article{wang_bulk_2024,
  title = {{Bulk HighTemperature Superconductivity in Pressurized Tetragonal La$_{2}$PrNi$_{2}$O$_{7}$}},
  author = {Wang, Ning-Ning and Wang, Gang and Shen, Xiao-Ling and Hou, Jun and Luo, Jun and Ma, Xiao-Ping and Yang, Huai-Xin and Shi, Li-Fen and Dou, Jie and Feng, Jie and Yang, Jie and Shi, Yun-Qing and Ren, Zhi-An and Ma, Han-Ming and Yang, Peng-Tao and Liu, Zi-Yi and Liu, Yue and Zhang, Hua and Dong, Xiao-Li and Wang, Yu-Xin and Jiang, Kun and Hu, Jiang-Ping and Nagasaki, Shoko and Kitagawa, Kentaro and Calder, Stuart and Yan, Jia-Qiang and Sun, Jian-Ping and Wang, Bo-Sen and Zhou, Rui and Uwatoko, Yoshiya and Cheng, Jin-Guang},
  year = 2024,
  month = oct,
  journal = {Nature},
  volume = {634},
  number = {8034},
  pages = {579--584},
  issn = {0028-0836, 1476-4687},
  doi = {10.1038/s41586-024-07996-8},
  url = {https://www.nature.com/articles/s41586-024-07996-8},
  urldate = {2026-03-10},
  langid = {english},
}

@article{wang_normal_2024,
  title = {{Normal and {{Superconducting Properties}} of La$_{3}$Ni$_{2}$O$_{7}$}},
  author = {Wang, Meng  and Wen , Hai-Hu  and Wu , Tao  and Yao , Dao-Xin  and Xiang , Tao },
  year = 2024,
  month = jul,
  journal = {Chin. Phys. Lett.},
  volume = {41},
  number = {7},
  pages = {077402},
  issn = {0256-307X, 1741-3540},
  doi = {10.1088/0256-307X/41/7/077402},
  url = {https://iopscience.iop.org/article/10.1088/0256-307X/41/7/077402},
  urldate = {2026-03-10},
  langid = {english},
  
}

@article{ko_signatures_2025a,
  title = {{Signatures of Ambient Pressure Superconductivity in Thin Film La$_{3}$Ni$_{2}$O$_{7}$}},
  author = {Ko, EunKyo and Yu, Yi-Jun and Liu, Yi-Di and Bhatt, Lopa and Li, Jia-Rui and Thampy, Vivek and Kuo, Cheng-Tai and Wang, Bai-Yang and Lee, Yong-Hun and Lee, Kyuho and Lee, Jun-Sik and Goodge, Berit H. and Muller, David A. and Hwang, Harold Y.},
  year = 2025,
  month = feb,
  journal = {Nature},
  volume = {638},
  number = {8052},
  pages = {935--940},
  issn = {0028-0836, 1476-4687},
  doi = {10.1038/s41586-024-08525-3},
  url = {https://www.nature.com/articles/s41586-024-08525-3},
  urldate = {2026-04-01},
  langid = {english},
  
}

@article{xiang_stabilizing_2026,
  title = {{Stabilizing and Tuning Superconductivity in La$_{3}$Ni$_{2}$O$_{7-\delta}$ Films: {{Oxygen}} Recycling Protocol Reveals HoleDoping Analogue}},
  author = {Xiang, Li-Fen and Lei, Si-Yi and Ren, Xiao-Lin and Han, Zi-Ao and Xu, Zi-Jian and Zhou, X. J. and Zhu, Zhi-Hai},
  year = 2026,
  month = mar,
  journal = {Phys. Rev. B},
  volume = {113},
  number = {10},
  pages = {104522},
  issn = {2469-9950, 2469-9969},
  doi = {10.1103/h9ls-y4s7},
  url = {https://link.aps.org/doi/10.1103/h9ls-y4s7},
  urldate = {2026-04-01},
  langid = {english},
  
}

@article{liu_superconductivity_2025,
  title = {{Superconductivity and NormalState Transport in Compressively Strained La$_{2}$PrNi$_{2}$O$_{7}$ Thin Films}},
  author = {Liu, Yi-Di and Ko, EunKyo and Tarn, Yao-Ju and Bhatt, Lopa and Li, Jia-Rui and Thampy, Vivek and Goodge, Berit H. and Muller, David A. and Raghu, Srinivas and Yu, Yi-Jun and Hwang, Harold Y.},
  year = 2025,
  month = aug,
  journal = {Nat. Mater.},
  volume = {24},
  number = {8},
  pages = {1221--1227},
  issn = {1476-1122, 1476-4660},
  doi = {10.1038/s41563-025-02258-y},
  url = {https://www.nature.com/articles/s41563-025-02258-y},
  urldate = {2025-08-19},
  langid = {english},
  
}

@article{zhou_ambientpressure_2025a,
  title = {{Ambient Pressure Superconductivity Onset above 40 {{K}} in (La,Pr)$_{3}$Ni$_{2}$O$_{7-\delta}$ Films}},
  author = {Zhou, Guang-Di and Lv, Wei and Wang, Heng and Nie, Zi-Hao and Chen, Ya-Qi and Li, Yue-Ying and Huang, Hao-Liang and Chen, Wei-Qiang and Sun, Yu-Jie and Xue, Qi-Kun and Chen, Zhuo-Yu},
  year = 2025,
  month = apr,
  journal = {Nature},
  volume = {640},
  number = {8059},
  pages = {641--646},
  issn = {0028-0836, 1476-4687},
  doi = {10.1038/s41586-025-08755-z},
  url = {https://www.nature.com/articles/s41586-025-08755-z},
  urldate = {2026-04-01},
  langid = {english},
 
}

@article{zhou_superconductivity_2026,
  title = {{Superconductivity Onset above 60 {{K}} in AmbientPressure Nickelate Films}},
  author = {Zhou, Guang-Di and Wang, Heng and Huang, Hao-Liang and Chen, Ya-Qi and Peng, Fei and Lv, Wei and Nie, Zi-Hao and Wang, Wei and Jia, Jin-Feng and Xue, Qi-Kun and Chen, Zhuo-Yu},
  year = 2026,
  month = mar,
  journal = {Natl. Sci. Rev.},
  pages = {nwag151},
  issn = {2095-5138, 2053-714X},
  doi = {10.1093/nsr/nwag151},
  url = {https://academic.oup.com/nsr/advance-article/doi/10.1093/nsr/nwag151/8512895},
  urldate = {2026-04-01},
  copyright = {https://creativecommons.org/licenses/by/4.0/},
  langid = {english},
  
}

@article{hao_superconductivity_2025,
  title = {{Superconductivity in {{Sr-doped La$_{3}$Ni$_{2}$O$_{7}$}} Thin Films}},
  author = {Hao, Bo and Wang, Mao-Sen and Sun, Wen-Jie and Yang, Yang and Mao, Zhang-Wen and Yan, Sheng-Jun and Sun, Hao-Ying and Zhang, Hong-Yi and Han, Lu and Gu, Zheng-Bin and Zhou, Jian and Ji, Dian-Xiang and Nie, Yue-Feng},
  year = 2025,
  month = nov,
  journal = {Nat. Mater.},
  volume = {24},
  number = {11},
  pages = {1756--1762},
  issn = {1476-1122, 1476-4660},
  doi = {10.1038/s41563-025-02327-2},
  url = {https://www.nature.com/articles/s41563-025-02327-2},
  urldate = {2026-04-01},
  langid = {english},
  
}

@article{ren_resolving_2025a,
  title = {{Resolving the Electronic Ground State of {{La$_3$Ni$_2$O$_{7-\delta}$}} Films}},
  author = {Ren, Xiao-Lin and Sutarto, Ronny and Wu, Xian-Xin and Zhang, Jian-Feng and Huang, Hai and Xiang, Tao and Hu, Jiang-Ping and Comin, Riccardo and Zhou, Xing-Jiang and Zhu, Zhi-Hai},
  year = 2025,
  month = feb,
  journal = {Commun. Phys.},
  volume = {8},
  number = {1},
  pages = {52},
  issn = {2399-3650},
  doi = {10.1038/s42005-025-01971-z},
  url = {https://www.nature.com/articles/s42005-025-01971-z},
  urldate = {2025-11-08},
  langid = {english}
}

@article{gupta_anisotropic_2025,
  title = {Anisotropic Spin Stripe Domains in Bilayer {{La$_3$Ni$_2$O$_7$}}},
  author = {Gupta, Naman K. and Gong, Rantong and Wu, Yi and Kang, Mingu and Parzyck, Christopher T. and Gregory, Benjamin Z. and Costa, Noah and Sutarto, Ronny and Sarker, Suchismita and Singer, Andrej and Schlom, Darrell G. and Shen, Kyle M. and Hawthorn, David G.},
  year = 2025,
  month = jul,
  journal = {Nature Communications},
  volume = {16},
  number = {1},
  pages = {6560},
  issn = {2041-1723},
  doi = {10.1038/s41467-025-61653-w},
  url = {https://www.nature.com/articles/s41467-025-61653-w},
  urldate = {2026-06-22},
  langid = {english}
}

@article{chen_3d_z2_2026,
  title = {{$3d_{z^2}$ Orbital Delocalization and Magnetic Collapse in Superconducting (La,Pr)$_{3}$Ni$_{2}$O$_{7-\delta}$ Films}},
  author = {Chen, Xiaoyang and Zhang, Wenliang and Peng, Fei and Cui, Ting and Zhou, Guangdi and Li, Zezhong and Choi, Jaewon and Xu, Lizhi and Chiu, Yiu-Fung and Agrestini, Stefano and Tippireddy, Sahil and Huang, Haoliang and Wang, Heng and Wu, Xianfeng and Li, Peng and Jia, Jin-Feng and {Garcia-Fernandez}, Mirian and Lu, Yi and Guo, Er-Jia and Xue, Qi-Kun and Chen, Zhuoyu and Feng, Donglai and Zhou, Ke-Jin},
  year = 2026,
  month = aug,
  journal = {National Science Review},
  pages = {nwag518},
  issn = {2095-5138, 2053-714X},
  doi = {10.1093/nsr/nwag518},
  url = {https://academic.oup.com/nsr/advance-article/doi/10.1093/nsr/nwag518/8770740},
  urldate = {2026-09-06},
  copyright = {https://creativecommons.org/licenses/by/4.0/},
  langid = {english}
}

@article{zhang_interlayer_2026,
  title = {Interlayer Hybridization Promotes Superconductivity in Bilayer Nickelates},
  author = {Zhang, Shilong and Zhang, Meng and Luo, Qiling and Tao, Zihao and Huang, Hsiao-Yu and Weschke, Eugen and Li, Kunhao and Channagowdra, Ganesha and Li, Jie and Fu, Junchi and Huang, Di-Jing and Xie, Yanwu and Lu, Yi and Peng, Yingying},
  year = 2026,
  month = sep,
  journal = {Nature Materials},
  issn = {1476-1122, 1476-4660},
  doi = {10.1038/s41563-026-02750-z},
  url = {https://www.nature.com/articles/s41563-026-02750-z},
  urldate = {2026-10-01},
  langid = {english}
}

@misc{li_interlayer_2026,
  title = {Interlayer {{Five-Spin Polaron}} in {{Superconducting Bilayer Nickelates}}},
  author = {Li, Jiarui and Parzyck, Christopher T. and Lomeli, Eder G. and Liu, Yidi and Kim, Taehun and Lee, Heemin and Zhuo, Zengqing and Ko, Eun Kyo and Tarn, Yaoju and Kuo, Cheng-Tai and Sutarto, Ronny and Jia, Chunjing and Thampy, Vivek and Pelliciari, Jonathan and Yang, Wanli and Moritz, Brian and Yu, Yijun and Lee, Jun-Sik and Bisogni, Valentina and Devereaux, Thomas P. and Hwang, Harold Y. and Lee, Wei-Sheng},
  year = 2026,
  month = jun,
  number = {arXiv:2605.02891},
  eprint = {2605.02891},
  primaryclass = {cond-mat.str-el},
  publisher = {arXiv},
  doi = {10.48550/ARXIV.2605.02891},
  url = {https://arxiv.org/abs/2605.02891},
  urldate = {2026-07-10},
  archiveprefix = {arXiv},
}

@misc{plokhikh_unraveling_2025,
  title = {{Unraveling Spin Density Wave Order in Layered Nickelates La$_{3}$Ni$_{2}$O$_{7}$ and La$_{2}$PrNi$_{2}$O$_{7}$ via Neutron Diffraction}},
  author = {Plokhikh, Igor and Hicken, Thomas J. and Keller, Lukas and Pomjakushin, Vladimir and Moody, Samuel H. and {Foury-Leylekian}, Pascale and Krieger, Jonas J. and Luetkens, Hubertus and Guguchia, Zurab and Khasanov, Rustem and Gawryluk, Dariusz Jakub},
  year = 2025,
  month = mar,
  number = {arXiv:2503.05287},
  eprint = {2503.05287},
  primaryclass = {cond-mat.supr-con},
  publisher = {arXiv},
  doi = {10.48550/arXiv.2503.05287},
  url = {http://arxiv.org/abs/2503.05287},
  urldate = {2026-06-22},
  archiveprefix = {arXiv},
  langid = {english}
}

@article{fradkin_colloquium_2015,
  title = {{\emph{Colloquium}} : {{Theory}} of Intertwined Orders in High Temperature Superconductors},
  shorttitle = {{\emph{Colloquium}}},
  author = {Fradkin, Eduardo and Kivelson, Steven A. and Tranquada, John M.},
  year = 2015,
  month = may,
  journal = {Reviews of Modern Physics},
  volume = {87},
  number = {2},
  pages = {457--482},
  issn = {0034-6861, 1539-0756},
  doi = {10.1103/RevModPhys.87.457},
  url = {https://link.aps.org/doi/10.1103/RevModPhys.87.457},
  urldate = {2026-07-11},
  copyright = {http://link.aps.org/licenses/aps-default-license},
  langid = {english}
}

@article{luo_bilayer_2023,
  title = {{Bilayer {{TwoOrbital Model}} of La$_{3}$Ni$_{2}$O$_{7}$ under {{Pressure}}}},
  author = {Luo, Zhi-Hui and Hu, Xun-Wu and Wang, Meng and W{\'u}, W{\'e}i and Yao, Dao-Xin},
  year = 2023,
  month = sep,
  journal = {Phys. Rev. Lett.},
  volume = {131},
  number = {12},
  pages = {126001},
  issn = {0031-9007, 1079-7114},
  doi = {10.1103/PhysRevLett.131.126001},
  url = {https://link.aps.org/doi/10.1103/PhysRevLett.131.126001},
  urldate = {2026-04-01},
  langid = {english}
}

@article{li_angleresolved_2025,
  title = {{AngleResolved Photoemission Spectroscopy of Superconducting (La,Pr)$_{3}$Ni$_{2}$O$_{7}$/{{SrLaAlO$_{4}$}} Heterostructures}},
  author = {Li, Peng and Zhou, Guang-Di and Lv, Wei and Li, Yue-Ying and Yue, Chang-Ming and Huang, Hao-Liang and Xu, Li-Zhi and Shen, Jian-Chang and Miao, Yu and Song, Wen-Hua and Nie, Zi-Hao and Chen, Ya-Qi and Wang, Heng and Chen, Wei-Qiang and Huang, Yao-Bo and Chen, Zhen-Hua and Qian, Tian and Lin, Jun-Hao and He, Jun-Feng and Sun, Yu-Jie and Chen, Zhuo-Yu and Xue, Qi-Kun},
  year = 2025,
  month = sep,
  journal = {Natl. Sci. Rev.},
  volume = {12},
  number = {10},
  pages = {nwaf205},
  issn = {2095-5138, 2053-714X},
  doi = {10.1093/nsr/nwaf205},
  url = {https://academic.oup.com/nsr/article/doi/10.1093/nsr/nwaf205/8140053},
  urldate = {2026-04-01},
  copyright = {https://creativecommons.org/licenses/by/4.0/},
  langid = {english},
  
}

@article{wang_electronic_2025,
  title = {Electronic structure of compressively strained bilayer nickelate thin film},
  author = {Wang, Bai Yang and Abadi, Sebastien N. and Liu, Yidi and Zhang, Yu and Zhong, Yong and Yu, Yijun and Goodge, Berit H. and Zhang, Xiaoliang and Wu, Yi-Ming and Wang, Ruohan and Li, Jiarui and Tarn, Yaoju and Ko, Eun Kyo and Thampy, Vivek and Lin, Chun and Hashimoto, Makoto and Lu, Donghui and Lee, Young S. and Devereaux, Thomas P. and Jia, Chunjing and Hwang, Harold Y. and Shen, Zhi-Xun},
  journal = {Phys. Rev. X},
  pages = {},
  year = {2026},
  month = {Jun},
  publisher = {American Physical Society},
  doi = {10.1103/4h5x-btvx},
  url = {https://link.aps.org/doi/10.1103/4h5x-btvx}
}

@article{liu_electronic_2024,
  title = {Electronic Correlations and Partial Gap in the Bilayer Nickelate {La$_{3}$Ni$_{2}$O$_{7}$}},
  author = {Liu, Zhe and Huo, Mengwu and Li, Jie and Li, Qing and Liu, Yuecong and Dai, Yaomin and Zhou, Xiaoxiang and Hao, Jiahao and Lu, Yi and Wang, Meng and Wen, Hai-Hu},
  year = 2024,
  month = aug,
  journal = {Nature Communications},
  volume = {15},
  number = {1},
  pages = {7570},
  issn = {2041-1723},
  doi = {10.1038/s41467-024-52001-5},
  url = {https://www.nature.com/articles/s41467-024-52001-5},
  urldate = {2026-07-11},
  langid = {english}
}

@article{yashima_microscopic_2025,
  title = {{Microscopic Evidence for Spin--Spinless Stripe Order with Reduced Ni Moments within \emph{ab} Plane for Bilayer Nickelate La$_{3}$Ni$_{2}$O$_{7}$ Probed by $^{139}$La-NQR}},
  author = {Yashima, Mitsuharu and Seto, Nina and Oshita, Yujiro and Kakoi, Masataka and Sakurai, Hiroya and Takano, Yoshihiko and Mukuda, Hidekazu},
  year = 2025,
  month = may,
  journal = {Journal of the Physical Society of Japan},
  volume = {94},
  number = {5},
  pages = {054704},
  issn = {0031-9015, 1347-4073},
  doi = {10.7566/JPSJ.94.054704},
  url = {https://journals.jps.jp/doi/10.7566/JPSJ.94.054704},
  urldate = {2026-07-11},
  langid = {english}
}

@article{chen_electronic_2024,
  title = {Electronic and Magnetic Excitations in {{La3Ni2O7}}},
  author = {Chen, Xiaoyang and Choi, Jaewon and Jiang, Zhicheng and Mei, Jiong and Jiang, Kun and Li, Jie and Agrestini, Stefano and {Garcia-Fernandez}, Mirian and Sun, Hualei and Huang, Xing and Shen, Dawei and Wang, Meng and Hu, Jiangping and Lu, Yi and Zhou, Ke-Jin and Feng, Donglai},
  year = 2024,
  month = nov,
  journal = {Nature Communications},
  volume = {15},
  number = {1},
  pages = {9597},
  issn = {2041-1723},
  doi = {10.1038/s41467-024-53863-5},
  url = {https://www.nature.com/articles/s41467-024-53863-5},
  urldate = {2026-06-26},
  langid = {english}
}

@article{nie_superconductivity_2026,
  title = {Superconductivity and Electronic Structures of Nickelate Thin Film Superstructures},
  author = {Nie, Zihao and Li, Yueying and Lv, Wei and Xu, Lizhi and Jiang, Zhicheng and Fu, Peng and Zhou, Guangdi and Song, Wenhua and Chen, Yaqi and Wang, Heng and Huang, Haoliang and Lin, Junhao and Jia, Jin-Feng and Shen, Dawei and Li, Peng and Xue, Qi-Kun and Chen, Zhuoyu},
  year = 2026,
  month = apr,
  journal = {Nature},
  volume = {652},
  number = {8110},
  pages = {628--634},
  issn = {0028-0836, 1476-4687},
  doi = {10.1038/s41586-026-10352-7},
  url = {https://www.nature.com/articles/s41586-026-10352-7},
  urldate = {2026-07-11},
  langid = {english}
}

@article{sun_observation_2025,
  title = {Observation of Superconductivity-Induced Leading-Edge Gap in a Bilayer Nickelate},
  author = {Sun, Wenjie and Jiang, Zhicheng and Hao, Bo and Yan, Shengjun and Zhang, Hongyi and Wang, Maosen and Yang, Yang and Sun, Haoying and Liu, Zhengtai and Ji, Dianxiang and Gu, Zhengbin and Zhou, Jian and Shen, Dawei and Feng, Donglai and Nie, Yuefeng},
  year = 2026,
  month = jul,
  journal = {Nature Physics},
  issn = {1745-2473, 1745-2481},
  doi = {10.1038/s41567-026-03396-z},
  url = {https://www.nature.com/articles/s41567-026-03396-z},
  urldate = {2026-07-22},
  langid = {english}
}

@article{zhang_structural_2024,
  title = {Structural Phase Transition, S\textpm -Wave Pairing, and Magnetic Stripe Order in Bilayered Superconductor {{La3Ni2O7}} under Pressure},
  author = {Zhang, Yang and Lin, Ling-Fang and Moreo, Adriana and Maier, Thomas A. and Dagotto, Elbio},
  year = 2024,
  month = mar,
  journal = {Nature Communications},
  volume = {15},
  number = {1},
  pages = {2470},
  issn = {2041-1723},
  doi = {10.1038/s41467-024-46622-z},
  url = {https://www.nature.com/articles/s41467-024-46622-z},
  urldate = {2026-07-11},
  langid = {english}
}

@article{ryee_superconductivity_2025,
  title = {Superconductivity {{Governed}} by {{Janus-Faced Fermiology}} in {{Strained Bilayer Nickelates}}},
  author = {Ryee, Siheon and Witt, Niklas and Sangiovanni, Giorgio and Wehling, Tim O.},
  year = 2025,
  month = dec,
  journal = {Physical Review Letters},
  volume = {135},
  number = {23},
  pages = {236003},
  issn = {0031-9007, 1079-7114},
  doi = {10.1103/ncbf-9b8m},
  url = {https://link.aps.org/doi/10.1103/ncbf-9b8m},
  urldate = {2026-07-11},
  langid = {english}
}

@article{zaanen_band_1985,
  title = {Band Gaps and Electronic Structure of Transition-Metal Compounds},
  author = {Zaanen, J. and Sawatzky, G. A. and Allen, J. W.},
  year = 1985,
  month = jul,
  journal = {Physical Review Letters},
  volume = {55},
  number = {4},
  pages = {418--421},
  issn = {0031-9007},
  doi = {10.1103/PhysRevLett.55.418},
  url = {https://link.aps.org/doi/10.1103/PhysRevLett.55.418},
  urldate = {2026-07-18},
  copyright = {http://link.aps.org/licenses/aps-default-license},
  langid = {english}
}

@article{liu_evidence_2023,
  title = {{Evidence for Charge and Spin Density Waves in Single Crystals of La$_{3}$Ni$_{2}$O$_{7}$ and La$_{3}$Ni$_{2}$O$_{6}$}},
  author = {Liu, Zengjia and Sun, Hualei and Huo, Mengwu and Ma, Xiaoyan and Ji, Yi and Yi, Enkui and Li, Lisi and Liu, Hui and Yu, Jia and Zhang, Ziyou and Chen, Zhiqiang and Liang, Feixiang and Dong, Hongliang and Guo, Hanjie and Zhong, Dingyong and Shen, Bing and Li, Shiliang and Wang, Meng},
  year = 2023,
  month = jan,
  journal = {Science China Physics, Mechanics \& Astronomy},
  volume = {66},
  number = {1},
  pages = {217411},
  issn = {1674-7348, 1869-1927},
  doi = {10.1007/s11433-022-1962-4},
  url = {https://link.springer.com/10.1007/s11433-022-1962-4},
  urldate = {2026-07-18},
  langid = {english}
}

@article{chen_evidence_2024,
  title = {{Evidence of Spin Density Waves in La$_{3}$Ni$_{2}$O$_{7-\delta}$}},
  author = {Chen, Kaiwen and Liu, Xiangqi and Jiao, Jiachen and Zou, Muyuan and Jiang, Chengyu and Li, Xin and Luo, Yixuan and Wu, Qiong and Zhang, Ningyuan and Guo, Yanfeng and Shu, Lei},
  year = 2024,
  month = jun,
  journal = {Physical Review Letters},
  volume = {132},
  number = {25},
  pages = {256503},
  issn = {0031-9007, 1079-7114},
  doi = {10.1103/PhysRevLett.132.256503},
  url = {https://link.aps.org/doi/10.1103/PhysRevLett.132.256503},
  urldate = {2026-07-18},
  langid = {english}
}

@article{qu_bilayer_2024,
  title = {{Bilayer $t$-$J$-$J_{\perp}$ Model and Magnetically Mediated Pairing in the Pressurized Nickelate La$_{3}$Ni$_{2}$O$_{7}$}},
  author = {Qu, Xing-Zhou and Qu, Dai-Wei and Chen, Jialin and Wu, Congjun and Yang, Fan and Li, Wei and Su, Gang},
  year = 2024,
  month = jan,
  journal = {Physical Review Letters},
  volume = {132},
  number = {3},
  pages = {036502},
  issn = {0031-9007, 1079-7114},
  doi = {10.1103/PhysRevLett.132.036502},
  url = {https://link.aps.org/doi/10.1103/PhysRevLett.132.036502},
  urldate = {2026-07-18},
  langid = {english}
}

@article{hawthorn_invacuum_2011,
  title = {An In-Vacuum Diffractometer for Resonant Elastic Soft x-Ray Scattering},
  author = {Hawthorn, D. G. and He, F. and Venema, L. and Davis, H. and Achkar, A. J. and Zhang, J. and Sutarto, R. and Wadati, H. and Radi, A. and Wilson, T. and Wright, G. and Shen, K. M. and Geck, J. and Zhang, H. and Nov{\'a}k, V. and Sawatzky, G. A.},
  year = 2011,
  month = jul,
  journal = {Review of Scientific Instruments},
  volume = {82},
  number = {7},
  pages = {073104},
  issn = {0034-6748, 1089-7623},
  doi = {10.1063/1.3607438},
  url = {https://pubs.aip.org/rsi/article/82/7/073104/354456/An-in-vacuum-diffractometer-for-resonant-elastic},
  urldate = {2023-11-13},
  langid = {english}
}

@article{pan_effect_2024,
  title = {{Effect of Rare-Earth Element Substitution in Superconducting $R_3$Ni$_2$O$_7$ under Pressure}},
  author = {Pan, Zhiming and Lu, Chen and Yang, Fan and Wu, Congjun},
  year = 2024,
  month = aug,
  journal = {Chinese Physics Letters},
  volume = {41},
  number = {8},
  pages = {087401},
  issn = {0256-307X, 1741-3540},
  doi = {10.1088/0256-307X/41/8/087401},
  url = {https://iopscience.iop.org/article/10.1088/0256-307X/41/8/087401},
  urldate = {2026-08-14}
}

@article{yang_orbitaldependent_2024,
  title = {{Orbital-dependent electron correlation in double-layer nickelate La$_3$Ni$_2$O$_7$}},
  author = {Yang, Jiangang and Sun, Hualei and Hu, Xunwu and Xie, Yuyang and Miao, Taimin and Luo, Hailan and Chen, Hao and Liang, Bo and Zhu, Wenpei and Qu, Gexing and Chen, Cui-Qun and Huo, Mengwu and Huang, Yaobo and Zhang, Shenjin and Zhang, Fengfeng and Yang, Feng and Wang, Zhimin and Peng, Qinjun and Mao, Hanqing and Liu, Guodong and Xu, Zuyan and Qian, Tian and Yao, Dao-Xin and Wang, Meng and Zhao, Lin and Zhou, X. J.},
  year = 2024,
  month = may,
  journal = {Nature Communications},
  volume = {15},
  number = {1},
  eprint = {2309.01148},
  primaryclass = {cond-mat.supr-con},
  pages = {4373},
  issn = {2041-1723},
  doi = {10.1038/s41467-024-48701-7},
  url = {http://arxiv.org/abs/2309.01148},
  urldate = {2026-06-22},
  archiveprefix = {arXiv},
  langid = {english}
}

@article{chang_direct_2012,
  title = {{Direct Observation of Competition between Superconductivity and Charge Density Wave Order in YBa$_2$Cu$_3$O$_{6.67}$}},
  author = {Chang, J. and Blackburn, E. and Holmes, A. T. and Christensen, N. B. and Larsen, J. and Mesot, J. and Liang, Ruixing and Bonn, D. A. and Hardy, W. N. and Watenphul, A. and Zimmermann, M. V. and Forgan, E. M. and Hayden, S. M.},
  year = 2012,
  month = dec,
  journal = {Nature Physics},
  volume = {8},
  number = {12},
  pages = {871--876},
  issn = {1745-2473, 1745-2481},
  doi = {10.1038/nphys2456},
  url = {https://www.nature.com/articles/nphys2456},
  urldate = {2026-09-04},
  langid = {english}
}

@article{ghiringhelli_longrange_2012,
  title = {{Long-Range Incommensurate Charge Fluctuations in (Y,Nd)Ba$_2$Cu$_3$O$_{6+x}$}},
  author = {Ghiringhelli, G. and Le Tacon, M. and Minola, M. and {Blanco-Canosa}, S. and Mazzoli, C. and Brookes, N. B. and De Luca, G. M. and Frano, A. and Hawthorn, D. G. and He, F. and Loew, T. and Sala, M. Moretti and Peets, D. C. and Salluzzo, M. and Schierle, E. and Sutarto, R. and Sawatzky, G. A. and Weschke, E. and Keimer, B. and Braicovich, L.},
  year = 2012,
  month = aug,
  journal = {Science},
  volume = {337},
  number = {6096},
  pages = {821--825},
  issn = {0036-8075, 1095-9203},
  doi = {10.1126/science.1223532},
  url = {https://www.science.org/doi/10.1126/science.1223532},
  urldate = {2026-09-04},
  langid = {english}
}

@article{khasanov_pressureenhanced_2025,
  title = {{Pressure-Enhanced Splitting of Density Wave Transitions in {{La$_3$Ni$_2$O$_{7-\delta}$}}}},
  author = {Khasanov, Rustem and Hicken, Thomas J. and Gawryluk, Dariusz J. and Sazgari, Vahid and Plokhikh, Igor and Sorel, Lo{\"\i}c Pierre and Bartkowiak, Marek and B{\"o}tzel, Steffen and Lechermann, Frank and Eremin, Ilya M. and Luetkens, Hubertus and Guguchia, Zurab},
  year = 2025,
  month = mar,
  journal = {Nature Physics},
  volume = {21},
  number = {3},
  pages = {430--436},
  issn = {1745-2473, 1745-2481},
  doi = {10.1038/s41567-024-02754-z},
  url = {https://www.nature.com/articles/s41567-024-02754-z},
  urldate = {2026-09-08},
  langid = {english}
}

@article{dong_visualization_2024,
  title = {Visualization of oxygen vacancies and self-doped ligand holes in {La$_3$Ni$_2$O$_{7-\delta}$}},
  author = {Dong, Zehao and Huo, Mengwu and Li, Jie and Li, Jingyuan and Li, Pengcheng and Sun, Hualei and Gu, Lin and Lu, Yi and Wang, Meng and Wang, Yayu and Chen, Zhen},
   year = 2024,
  month = jun,
  journal = {Nature},
  volume = {630},
  number = {8018},
  pages = {847--852},
  issn = {0028-0836, 1476-4687},
  doi = {10.1038/s41586-024-07482-1},
  url = {https://www.nature.com/articles/s41586-024-07482-1},
  urldate = {2026-07-18},
  langid = {english}
}

@article{dong_interstitial_2025,
  title = {Interstitial oxygen order and its competition with superconductivity in {La$_2$PrNi$_2$O$_{7+\delta}$}},
  author = {Dong, Zehao and Wang, Gang and Wang, Ningning and Dong, Wen-Han and Gu, Lin and Xu, Yong and Cheng, Jinguang and Chen, Zhen and Wang, Yayu},
  year = 2025,
  month = dec,
  journal = {Nature Materials},
  volume = {24},
  number = {12},
  pages = {1927--1934},
  issn = {1476-1122, 1476-4660},
  doi = {10.1038/s41563-025-02351-2},
  url = {https://www.nature.com/articles/s41563-025-02351-2},
  urldate = {2026-09-22},
  langid = {english}
}

@misc{chen_nature_2026,
  title = {Nature of magnetism in bilayer nickelate {La$_3$Ni$_2$O$_7$} single crystals},
  author = {Chen, Lixing and Zhang, Enkang and Hao, Yiqing and Zhu, Yinghao and Cui, Bingkun and Abernathy, Douglas L. and Williams, Travis J. and Ikeda, Yoichi and Zhang, Hao and Liu, Feiyang and Wang, Wen-biao and Wang, Qisi and Zhao, Jun},
   year={2026},
    eprint={2605.03448},
    archivePrefix={arXiv},
    primaryClass={cond-mat.str-el},
    url={https://arxiv.org/abs/2605.03448}, 
}

@article{qiu_hightemperature_2026,
  title = {{High-Temperature Superconductivity at 100\,K in La$_{3-x}$Nd$_x$Ni$_2$O$_7$}},
  author = {Qiu, Zhengyang and Chen, Junfeng and Semenok, Dmitrii V. and Zhong, Qingyi and Zhou, Di and Li, Jingyuan and Ma, Peiyue and Huang, Xing and Huo, Mengwu and Xie, Tao and Chen, Xiang and Mao, Ho-kwang and Struzhkin, Viktor V. and Sun, Hualei and Wang, Meng},
  year = 2026,
  month = aug,
  journal = {Nature Communications},
  volume = {17},
  number = {1},
  pages = {9675},
  issn = {2041-1723},
  doi = {10.1038/s41467-026-76534-z},
  url = {https://www.nature.com/articles/s41467-026-76534-z},
  urldate = {2026-09-13},
  langid = {english}
}

@article{lechermann_electronic_2023,
  title = {{Electronic Correlations and Superconducting Instability in La$_3$Ni$_2$O$_7$ under High Pressure}},
  author = {Lechermann, Frank and Gondolf, Jannik and B{\"o}tzel, Steffen and Eremin, Ilya M.},
  year = 2023,
  month = nov,
  journal = {Physical Review B},
  volume = {108},
  number = {20},
  pages = {L201121},
  issn = {2469-9950, 2469-9969},
  doi = {10.1103/PhysRevB.108.L201121},
  url = {https://link.aps.org/doi/10.1103/PhysRevB.108.L201121},
  urldate = {2026-09-13},
  langid = {english}
}

@article{christiansson_correlated_2023,
  title = {{Correlated Electronic Structure of La$_3$Ni$_2$O$_7$ under Pressure}},
  author = {Christiansson, Viktor and Petocchi, Francesco and Werner, Philipp},
  year = 2023,
  month = nov,
  journal = {Physical Review Letters},
  volume = {131},
  number = {20},
  pages = {206501},
  issn = {0031-9007, 1079-7114},
  doi = {10.1103/PhysRevLett.131.206501},
  url = {https://link.aps.org/doi/10.1103/PhysRevLett.131.206501},
  urldate = {2026-09-13},
  langid = {english}
}

@article{shilenko_correlated_2023,
  title = {{Correlated Electronic Structure, Orbital-Selective Behavior, and Magnetic Correlations in Double-Layer La$_3$Ni$_2$O$_7$ under Pressure}},
  author = {Shilenko, D. A. and Leonov, I. V.},
  year = 2023,
  month = sep,
  journal = {Physical Review B},
  volume = {108},
  number = {12},
  pages = {125105},
  issn = {2469-9950, 2469-9969},
  doi = {10.1103/PhysRevB.108.125105},
  url = {https://link.aps.org/doi/10.1103/PhysRevB.108.125105},
  urldate = {2026-09-13},
  langid = {english}
}

@article{yang_interlayer_2023,
  title = {{Interlayer Valence Bonds and Two-Component Theory for High-$T_c$ Superconductivity of La$_3$Ni$_2$O$_7$ under Pressure}},
  author = {Yang, Yi-feng and Zhang, Guang-Ming and Zhang, Fu-Chun},
  year = 2023,
  month = nov,
  journal = {Physical Review B},
  volume = {108},
  number = {20},
  pages = {L201108},
  issn = {2469-9950, 2469-9969},
  doi = {10.1103/PhysRevB.108.L201108},
  url = {https://link.aps.org/doi/10.1103/PhysRevB.108.L201108},
  urldate = {2026-09-13},
  langid = {english}
}

@article{jiang_hightemperature_2024,
  title = {{High-Temperature Superconductivity in La$_3$Ni$_2$O$_7$}},
  author = {Jiang, Kun and Wang, Zi-Qiang and Zhang, Fu-Chun},
  year = 2024,
  month = jan,
  journal = {Chinese Physics Letters},
  volume = {41},
  number = {1},
  pages = {017402},
  issn = {0256-307X, 1741-3540},
  doi = {10.1088/0256-307X/41/1/017402},
  url = {https://iopscience.iop.org/article/10.1088/0256-307X/41/1/017402},
  urldate = {2026-09-13},
  langid = {english}
}

@article{sakakibara_possible_2024,
  title = {{Possible High $T_c$ Superconductivity in La$_3$Ni$_2$O$_7$ under High Pressure through Manifestation of a Nearly Half-Filled Bilayer Hubbard Model}},
  author = {Sakakibara, Hirofumi and Kitamine, Naoya and Ochi, Masayuki and Kuroki, Kazuhiko},
  year = 2024,
  month = mar,
  journal = {Physical Review Letters},
  volume = {132},
  number = {10},
  pages = {106002},
  issn = {0031-9007, 1079-7114},
  doi = {10.1103/PhysRevLett.132.106002},
  url = {https://link.aps.org/doi/10.1103/PhysRevLett.132.106002},
  urldate = {2026-09-13},
  langid = {english}
}

@misc{mei_itinerant_2026,
  title = {{Itinerant Nature of Spin-Density-Wave Order in Ruddlesden-Popper Nickelates}},
  author = {Mei, Jiong and Xie, Tianyang and Jiang, Kun},
  year = {2026},
  month = {may},
  number = {arXiv:2605.20148},
  eprint = {2605.20148},
  primaryclass = {cond-mat.supr-con},
  publisher = {arXiv},
  doi = {10.48550/arXiv.2605.20148},
  url = {https://arxiv.org/abs/2605.20148},
  archiveprefix = {arXiv},
}
\end{document}